# Dirac and Klein-Gordon equations with equal scalar and vector potentials

A. D. Alhaidari[a,1], H. Bahlouli[a,2], A. Al-Hasan[b,3]

[a] *Physics Department, King Fahd University of Petroleum & Minerals, Dhahran 31261, Saudi Arabia*
[b] *Physics Department, King Faisal University, Al-Ahsa 31982, Saudi Arabia*

We study the three-dimensional Dirac and Klein-Gordon equations with scalar and vector potentials of equal magnitudes as an attempt to give a proper physical interpretation of this class of problems which has recently been accumulating interest. We consider a large class of these problems in which the potentials are noncentral (angular-dependent) such that the equations separate completely in spherical coordinates. The relativistic energy spectra are obtained and shown to differ from those of well-known problems that have the same nonrelativistic limit. Consequently, such problems should not be misinterpreted as the relativistic extension of the given potentials despite the fact that the nonrelativistic limit is the same. The Coulomb, Oscillator and Hartmann potentials are considered. This shows that although the nonrelativistic limit is well-defined and unique, the relativistic extension is not. Additionally, we investigate the Klein-Gordon equation with uneven mix of potentials leading to the correct relativistic extension. We consider the case of spherically symmetric exponential-type potentials resulting in the s-wave Klein-Gordon-Morse problem.

PACS numbers: 03.65.Ge, 03.65.Pm, 03.65.Ca

Keywords: Dirac equation, Klein-Gordon equation, scalar and vector coupling, noncentral potentials, relativistic extensions, Hartmann potential, Morse potential, pseudo-spin symmetry.

## I. INTRODUCTION

The success of quantum mechanics in the description of the atomic and sub-micro world is very impressive and overwhelming. Supplementing this theory with special relativity created one of the most accurate physical theories in recent history. An example is quantum electrodynamics; the theory that describes the interaction of charged particles with the electromagnetic radiation at high speeds or strong coupling. The Dirac equation and the Klein-Gordon (K-G) equation are the most frequently used wave equations for the description of particle dynamics in relativistic quantum mechanics. The fact that these two equations, for free particles, are constructed using two objects: the four-vector linear momentum operator $P_\mu = i\hbar \partial_\mu$ and the scalar rest-mass m, allows one to introduce naturally two types of potential coupling. One is the gauge invariant coupling to the four-vector potential $\{A_\mu(t,\vec{r})\}_{\mu=0}^{3}$ which is introduced via the minimal substitution $P_\mu \rightarrow P_\mu - gA_\mu$, where $g$ is a real coupling parameter. The other, is an additional coupling to the space-time scalar potential $S(t,\vec{r})$ which is introduced by the substitution $m \rightarrow m + S$. The term "four-vector" and "scalar" refers to the corresponding unitary irreducible representation of the Poincaré space-time symmetry group (the group of rotations and translations in 3+1 dimensional Minkowski space-time). Gauge invariance of the vector coupling allows for the freedom to fix the gauge (eliminate the nonphysical gauge modes) without altering the physical content of the problem. There are many choices of gauge fixing that one could impose. The Lorentz gauge, $\partial \cdot A = 0$, and the Coulomb gauge, $\vec{\nabla} \cdot \vec{A} = 0$, are two of the most commonly used conditions. However, many choose to simplify the solution of the problem by taking the space component of the vector potential to vanish (i.e., $\vec{A} = 0$). If we adapt this later choice and write the time component of the four-vector potential as $gA_0 = V(t, \vec{r})$, then we end up with two independent potential functions in the Dirac and K-G equations. These are the "vector" potential $V$ and the scalar potential $S$.

---
[1] E-mail: haidari@mailaps.org (corresponding author)
[2] E-mail: bahlouli@kfupm.edu.sa
[3] E-mail: aawahab76@yahoo.com

In the relativistic units, $\hbar = c = 1$, the free Dirac and K-G equations are written as $(i\gamma^\mu \partial_\mu - m)\Psi_D(t, \vec{r}) = 0$ and $(\partial^\mu \partial_\mu + m^2)\Psi_{KG}(t, \vec{r}) = 0$, respectively. The convention of summing over repeated indices is used. For particles of spin $\frac{1}{2}$, $\{\gamma^\mu\}$ are 4×4 constant matrices with the following standard representation

$$\gamma^0 = \begin{pmatrix} I & 0 \\ 0 & -I \end{pmatrix}, \quad \vec{\gamma} = \begin{pmatrix} 0 & \vec{\sigma} \\ -\vec{\sigma} & 0 \end{pmatrix}, \tag{1.1}$$

where $I$ is the 2×2 unit matrix and $\vec{\sigma}$ are the three 2×2 hermitian Pauli spin matrices. The vector and scalar couplings mentioned above introduce potential interactions by mapping the free Dirac and K-G equations above into the following

$$\left\{\gamma^0\left[i\frac{\partial}{\partial t} - V(t, \vec{r})\right] + i\vec{\gamma} \cdot \vec{\nabla} - m - S(t, \vec{r})\right\} \Psi_D(t, \vec{r}) = 0, \tag{1.2a}$$

$$\left\{-\left[i\frac{\partial}{\partial t} - V(\vec{r})\right]^2 - \vec{\nabla}^2 + [S(\vec{r}) + m]^2\right\} \psi_{KG}(\vec{r}) = 0, \tag{1.2b}$$

respectively. Recently, interest in the solutions of these two equations for the case where $S(\vec{r}) = \pm V(\vec{r})$ has surged. For the most recent contributions, with citations to earlier work, one may consult the papers in [1] and references therein.

In this article, we set out to present a critical investigation of this case, $S = \pm V$, considering the more general situation where the potentials are angular-dependent (noncentral) such that the Dirac and K-G equations are completely separable in spherical coordinates. Our targets are studies that present an improper physical interpretation of the solutions of such problems. However, studies that are aimed at the investigation of issues that are relevant to these problems such as the pseudo-spin symmetry in nuclear physics [2] and the effect of the breaking of this symmetry do not fall within the scope of our work. We will show (by example) that the solutions of such problems do not coincide with those of well-established problems that have the same nonrelativistic limit. The Coulomb, Oscillator and Hartmann potentials will be studied. This constitutes a (sufficient) proof that although the nonrelativistic limit is well-defined and unique, the relativistic extension is not. Thus, the physical interpretation of the relativistic problem should not be confused with those that may have the same nonrelativistic spectrum or similar structure of the potential functions. The formulation of the problem will be carried out in the following section whereas illuminating solutions for several examples will be obtained in Sec. III. The K-G equation with $S = \eta V$ will be investigated in Sec. IV, where $\eta$ is a real parameter such that $\eta \neq \pm 1$. We consider the example of radial exponential potentials and show that in this case the correct relativistic description (the S-wave Klein-Gordon-Morse problem) is obtained as long as $\eta$ relates to the physical parameters of the problem in a special way and such that $|\eta| > 1$.

## II. FORMULATION OF THE PROBLEM AND SOLUTION OF THE ANGULAR EQUATIONS

For time-independent potentials we can write the total wavefunction as $\Psi(t,\vec{r}) = e^{-i\varepsilon t}\psi(\vec{r})$, where $\varepsilon$ is the relativistic energy. Consequently, the Dirac and K-G equations (1.2a) and (1.2b) become, respectively

$$\begin{pmatrix} m + S(\vec{r}) + V(\vec{r}) & -i\vec{\sigma}\cdot\vec{\nabla} \\ -i\vec{\sigma}\cdot\vec{\nabla} & -m - S(\vec{r}) + V(\vec{r}) \end{pmatrix} \begin{pmatrix} \psi_+(\vec{r}) \\ \psi_-(\vec{r}) \end{pmatrix} = \varepsilon \begin{pmatrix} \psi_+(\vec{r}) \\ \psi_-(\vec{r}) \end{pmatrix}, \tag{2.1a}$$

$$\left\{ \vec{\nabla}^2 + [V(\vec{r}) - \varepsilon]^2 - [S(\vec{r}) + m]^2 \right\} \psi_{KG}(\vec{r}) = 0, \tag{2.1b}$$

where $\psi_+$ ($\psi_-$) is the top (bottom) two-component spinor of $\psi_D$. Now, if we take $S = \pm V$, then the potential contribution in the Dirac Hamiltonian will be either $\begin{pmatrix} 2V & 0 \\ 0 & 0 \end{pmatrix}$ or $\begin{pmatrix} 0 & 0 \\ 0 & 2V \end{pmatrix}$. It is believed that such a potential structure might result in irregular behavior of the solution. Nevertheless, with $S = \pm V$ equation (2.1a) gives one spinor component in terms of the other as

$$\psi_{\mp}(\vec{r}) = \frac{1}{\varepsilon \pm m}\left(-i\vec{\sigma}\cdot\vec{\nabla}\right)\psi_{\pm}(\vec{r}), \tag{2.2}$$

where $\varepsilon \neq \mp m$. This equation is referred to as the "kinetic balance" relation. Since $\varepsilon = +m$ ($\varepsilon = -m$) is an element of the positive (negative) energy spectrum of the Dirac Hamiltonian, then this relation with the top (bottom) sign is *not* valid for the negative (positive) energy solutions. Substituting from Eq. (2.2) into the Dirac equation (2.1a), with $S = \pm V$, results in the following second order differential equation

$$\left[ \vec{\nabla}^2 - 2(\varepsilon \pm m)V(\vec{r}) + \varepsilon^2 - m^2 \right]\psi_{\pm}(\vec{r}) = 0, \tag{2.3}$$

giving $\psi_+$ ($\psi_-$) as an element of the positive (negative) energy solutions. To obtain the other spinor component, we use the kinetic balance relation (2.2) with the top (bottom) sign. Therefore, the choice $S = +V$ ($S = -V$) dictates that the solution of Eq. (2.3) does not include the negative (positive) energy states. This observation highlights the second critical property in this kind of problems that has to be considered carefully when presenting the physical interpretation. It associates with each choice of potential configuration one sector of the energy spectrum, only the positive or the negative, but not both. This unsymmetrical treatment of the energy spectrum, where half of the spectrum is missing, is known to create problems such as particle-antiparticle interpretation of the relativistic theory [3]. As for the K-G equation (2.1b), we obtain the following, for $S = \pm V$

$$\left[ \vec{\nabla}^2 - 2(\varepsilon \pm m)V(\vec{r}) + \varepsilon^2 - m^2 \right]\psi_{KG}(\vec{r}) = 0, \tag{2.4}$$

which is identical to Eq. (2.3) for $\psi_{\pm}$. This equivalence of the Dirac representation to the K-G representation *in the presence of interaction* constitutes a constraint on the physical interpretation since it diminishes the well-known advantages of the former over the latter [3]. Moreover, the nonrelativistic limit, which is obtained by taking $\varepsilon - m \cong E$ where $|E| \ll m$, implies that the negative energy solutions (corresponding to $S = -V$) are free fields since under these conditions Eq. (2.3) and Eq. (2.4) reduce to

$$\left( \frac{1}{2m}\vec{\nabla}^2 + E \right)\psi(\vec{r}) = 0, \tag{2.5}$$

where $E$ is the nonrelativistic energy and $\psi$ stands for $\psi_-$ or $\psi_{KG}$. On the other hand, the positive energy states (where $S = +V$) in the nonrelativistic limit are solutions of

$$\left[\tfrac{1}{2m}\vec{\nabla}^2 - 2V(\vec{r}) + E\right]\psi(\vec{r}) = 0, \tag{2.6}$$

where $\psi$ stands for either $\psi_+$ or $\psi_{KG}$. This is the Schrödinger equation for the potential $2V$. Thus, we conclude that only the choice $S = +V$ produces a nontrivial nonrelativistic limit with a potential function $2V$, and not $V$. Accordingly, it would be natural to scale the potential terms in Eq. (2.1a) and Eq. (2.1b) so that in the nonrelativistic limit the interaction potential becomes $V$, not $2V$. Therefore, we modify Eq. (2.1a) and Eq. (2.1b) to read as follows:

$$\begin{pmatrix} m + \frac{V(\vec{r}) + S(\vec{r})}{2} & -i\vec{\sigma}\cdot\vec{\nabla} \\ -i\vec{\sigma}\cdot\vec{\nabla} & -m + \frac{V(\vec{r}) - S(\vec{r})}{2} \end{pmatrix} \begin{pmatrix} \psi_+(\vec{r}) \\ \psi_-(\vec{r}) \end{pmatrix} = \varepsilon \begin{pmatrix} \psi_+(\vec{r}) \\ \psi_-(\vec{r}) \end{pmatrix}, \tag{2.1a}'$$

$$\left\{\vec{\nabla}^2 + \left[\tfrac{1}{2}V(\vec{r}) - \varepsilon\right]^2 - \left[\tfrac{1}{2}S(\vec{r}) + m\right]^2\right\}\psi_{KG}(\vec{r}) = 0. \tag{2.1b}'$$

Consequently, Eqs. (2.3), (2.4) and (2.6) become

$$\left[\vec{\nabla}^2 - (\varepsilon \pm m)V(\vec{r}) + \varepsilon^2 - m^2\right]\psi_\pm(\vec{r}) = 0, \tag{2.3}'$$

$$\left[\vec{\nabla}^2 - (\varepsilon \pm m)V(\vec{r}) + \varepsilon^2 - m^2\right]\psi_{KG}(\vec{r}) = 0, \tag{2.4}'$$

$$\left[\tfrac{1}{2m}\vec{\nabla}^2 - V(\vec{r}) + E\right]\psi_+(\vec{r}) = 0, \tag{2.6}'$$

respectively. We are unable to make any further general statements beyond the three observations made above: (i) the singular matrix structure of the potential, (ii) the unsymmetrical treatment of the positive and negative energy spectrum, and (iii) the unfavorable equivalence of the Dirac equation (2.3) to the K-G equation (2.4) in the presence of interaction. Therefore, we adapt an alternative investigation strategy based on "demonstration by example". In other words, we make several choices of potential functions $V(\vec{r})$ that have well established relativistic extensions (e.g., the Dirac-Coulomb problem for which $V \sim \frac{1}{r}$) and compare their positive energy solutions with those obtained from Eq. (2.3)' or, equivalently, Eq. (2.4)' for $S = +V$. We choose not to pursue the case $S = -V$ since its nonrelativistic limit is the trivial interaction-free mode. This, of course, does not diminish the importance of such problems. It only limits its contribution (with the proper physical interpretation) to the relativistic regime.

Equation (2.3)' for $\psi_+$ and Eq. (2.4)' for $\psi_{KG}$ with $S = +V$ and for a general noncentral potential $V(r,\theta,\phi)$ could be written in spherical coordinates as follows

$$\left\{\frac{1}{r^2}\frac{\partial}{\partial r}r^2\frac{\partial}{\partial r} + \frac{1}{r^2}\left[(1-x^2)\frac{\partial^2}{\partial x^2} - 2x\frac{\partial}{\partial x} + \frac{1}{1-x^2}\frac{\partial^2}{\partial \phi^2}\right]\right.$$
$$\left. -(\varepsilon+m)V + \varepsilon^2 - m^2\right\}\psi = 0, \tag{2.7}$$

where $x = \cos\theta$ and $\psi$ stands for either $\psi_+$ or $\psi_{KG}$. Consequently, this equation is completely separable for potentials of the form

$$V(\vec{r}) = V_r(r) + \frac{1}{r^2}\left[V_\theta(x) + \frac{1}{1-x^2}V_\phi(\phi)\right]. \tag{2.8}$$

This is so because if we write the total wavefunction as $\psi(r,\theta,\phi) = r^{-1}R(r)\Theta(\theta)\Phi(\phi)$, then the wave equation (2.7) with the potential (2.8) gets completely separated in all three coordinates and as follows

$$\left[\frac{d^2}{d\phi^2}-(\varepsilon+\mathrm{m})V_\phi+E_\phi\right]\Phi=0, \tag{2.9a}$$

$$\left[(1-x^2)\frac{d^2}{dx^2}-2x\frac{d}{dx}-\frac{E_\phi}{1-x^2}-(\varepsilon+\mathrm{m})V_\theta+E_\theta\right]\Theta=0, \tag{2.9b}$$

$$\left[\frac{d^2}{dr^2}-\frac{E_\theta}{r^2}-(\varepsilon+\mathrm{m})V_r+\varepsilon^2-\mathrm{m}^2\right]R=0, \tag{2.9c}$$

where $E_\phi$ and $E_\theta$ are the separation constants, which are real and dimensionless. The components of the wavefunction are required to satisfy the boundary conditions. That is, $R(0)=R(\infty)=0$, $\Phi(\phi)=\Phi(\phi+2\pi)$, $\Theta(0)$ and $\Theta(\pi)$ are finite. If we specialize to the case where $V_\phi=0$, then the normalized solution of Eq. (2.9a) that satisfies the boundary conditions is

$$\Phi_m(\phi)=\frac{1}{\sqrt{2\pi}}e^{im\phi}, \quad m\in\mathbb{Z}=0,\pm1,\pm2,.., \tag{2.10}$$

giving $E_\phi=m^2$. The italic letter *m*, which stands for integers, should not be confused with the letter m that refers to the rest mass of the particle.

The solution of Eq. (2.9b) for bound states will be spanned by $L^2$ functions that are defined in the compact space with coordinate $x\in[-1,+1]$. Comparing it to the differential equation of the Jacobi polynomial $P_n^{(\mu,\nu)}(x)$ [4], which is also defined in the same space, we could suggest the following form of solution

$$\Theta(\theta)=A(1-x)^\alpha(1+x)^\beta P_n^{(\mu,\nu)}(x), \tag{2.11}$$

where *A* is the normalization constant. The real dimensionless parameters $\mu$ and $\nu$ are such that $\mu>-1$ and $\nu>-1$. Square integrability and the boundary conditions require that $\alpha>0$ and $\beta>0$. Substituting (2.11) into Eq. (2.9b) and using the differential equation of the Jacobi polynomials we obtain

$$\left\{\left[\mu-\nu-2\alpha+2\beta+x(\mu+\nu-2\alpha-2\beta)\right]\frac{d}{dx}+2x\left(\frac{\alpha}{1-x}-\frac{\beta}{1+x}\right)+\alpha(\alpha-1)\frac{1+x}{1-x}\right.$$
$$\left.+\beta(\beta-1)\frac{1-x}{1+x}-\frac{E_\phi}{1-x^2}-(\varepsilon+\mathrm{m})V_\theta+E_\theta-2\alpha\beta-n(n+\mu+\nu+1)\right\}P_n^{(\mu,\nu)}=0. \tag{2.12}$$

Requiring that the representation in the solution space, which is spanned by (2.11), be orthogonal dictates that the *x*-dependent factors multiplying $P_n^{(\mu,\nu)}$ and $\frac{d}{dx}P_n^{(\mu,\nu)}$ in Eq. (2.12) must vanish. Thus, the angular potential function $V_\theta(x)$ should be of the following form

$$V_\theta(x)=\frac{a+bx}{1-x^2}, \tag{2.13}$$

where *a* and *b* are real parameters. Additionally, simple manipulations of Eq. (2.12) with this potential function give the following results:

$$\mu=\sqrt{m^2+(\varepsilon+\mathrm{m})(a+b)}, \tag{2.14a}$$

$$\nu=\sqrt{m^2+(\varepsilon+\mathrm{m})(a-b)}, \tag{2.14b}$$

$$\alpha=\mu/2, \quad \beta=\nu/2, \tag{2.14c}$$

$$E_\theta = \left(n + \alpha + \beta + \tfrac{1}{2}\right)^2 - \tfrac{1}{4}. \tag{2.14d}$$

The Aharonov-Bohm [5] and Hartmann [6] potentials are special cases of (2.13) for which $b = 0$. For pure Aharonov-Bohm effect, $a$ is discrete via its linear dependence on the integer $m$. On the other hand, for the Hartmann problem the angular potential (2.13) should be supplemented by the radial Coulomb potential. The case where $b = \pm a$ corresponds to the magnetic monopole potential with singularity along the $\pm z$ axis [7]. The orthogonality relation of the Jacobi polynomials gives the following expression for the normalization constant that makes the angular wavefunctions $\{\Theta_n(\theta)\}$ an orthonormal set

$$A = \sqrt{\frac{2n+\mu+\nu+1}{2^{\mu+\nu+1}} \frac{\Gamma(n+1)\Gamma(n+\mu+\nu+1)}{\Gamma(n+\mu+1)\Gamma(n+\nu+1)}}. \tag{2.15}$$

Equation (2.14d) implies that for real representations we can always write $E_\theta = \lambda(\lambda+1)$, where $\lambda$ is a real number, not necessarily an integer but discrete (i.e., numerable). Additionally, $\lambda$ is evaluated as

$$\lambda = \begin{cases} n+\alpha+\beta & ,\lambda > 0 \\ -(n+\alpha+\beta)-1 & ,\lambda < -1 \end{cases} \tag{2.16}$$

However, real representations require that the expressions under the two square roots in Eq. (2.14a) and Eq. (2.14b) are non-negative. In other words, the absolute value of the integer $m$ should be not less than some positive integer $M$, where $M$ is the minimum integer that satisfies the following

$$M^2 \geq (\varepsilon + \mathrm{m})(|b| - a), \tag{2.17}$$

where we have assumed positive energy, corresponding to $S = +V$. Thus, the range of the integer $m$ becomes $m = \pm M, \pm(M+1), \pm(M+2),...$ and $M = 0$ only if $a \geq |b|$. For a given integer $m$ in this range and for a given physical parameter $\lambda$, the integer $n$ is not arbitrary but is determined by Eq. (2.16). Finally, we can write the complete orthonormal angular wavefunction as

$$\Omega_{nm}(\theta,\phi) = \sqrt{\frac{2n+\mu+\nu+1}{4\pi \times 2^{\mu+\nu}} \frac{\Gamma(n+1)\Gamma(n+\mu+\nu+1)}{\Gamma(n+\mu+1)\Gamma(n+\nu+1)}} (1-x)^{\frac{\mu}{2}}(1+x)^{\frac{\nu}{2}} P_n^{(\mu,\nu)}(x) \times e^{im\phi}, \tag{2.18}$$

where the dependence on $m$ comes also from the parameters $\mu$ and $\nu$ as given by Eq. (2.14a) and Eq. (2.14b), respectively.

In the following section, we solve the radial equation and obtain the energy spectra for several radial potential functions, $V_r(r)$. These will be compared with well established results in order to make general judgment and conclusions about the validity of the interpretation of the solutions obtained for $S = \pm V$.

### III. SOLUTIONS OF THE RADIAL EQUATION AND ENERGY SPECTRA FOR SEVERAL EXAMPLES

In this section we solve the radial equation, Eq. (2.9c), and obtain the positive energy spectra for the Coulomb, Oscillator and Hartmann potentials. We compare the results obtained in the first two cases with those of the well-known Dirac-Coulomb and Dirac-Oscillator problems that have the same nonrelativistic spectra. However, in the absence of a known exact relativistic extension of the Hartmann potential, we investigate the results of this case only under certain conditions. We limit our investigation to the energy spectrum since it is sufficient (for the purpose of our study) to reach definitive conclusions.

For bound states, the solution of the radial equation, Eq. (2.9c), will be spanned by $L^2$ functions defined on the positive real line with coordinate $y \in \Re^+$, where $y$ is proportional to some power of $r$. The following ansatz is compatible with this requirement and can be made to satisfy the boundary conditions

$$R(r) = B\, y^\tau e^{-\xi y}\, {}_1F_1(p;q;y), \qquad (3.1)$$

where $B$ is the normalization constant, ${}_1F_1$ is the confluent hypergeometric series [4], $p$ and $q$ are dimensionless real parameters. Square integrability and the boundary conditions require that the real dimensionless parameters $\tau$ and $\xi$ be positive. We will consider two cases. One, in which $y = \rho r$ and in the other $y = (\rho r)^2$, where $\rho$ is a positive real parameter of inverse length dimension. Substituting (3.1) into Eq. (2.9c) for $y = \rho r$, $E_\theta = \lambda(\lambda+1)$ and using the differential equation of the confluent hypergeometric series we obtain

$$\left[\left(1 - 2\xi + \frac{2\tau - q}{y}\right)\frac{d}{dy} + \frac{\tau(\tau-1) - \lambda(\lambda+1)}{y^2} + \frac{p - 2\tau\xi}{y} - \frac{\varepsilon + \mathrm{m}}{\rho^2}V_r + \xi^2 + \frac{\varepsilon^2 - \mathrm{m}^2}{\rho^2}\right]{}_1F_1 = 0. \qquad (3.2)$$

Requiring orthogonal representation for the bound states dictates that the $y$-dependent factors multiplying ${}_1F_1$ and $\frac{d}{dy}{}_1F_1$ in Eq. (3.2) must independently vanish. Consequently, the radial potential function should be proportional to $\frac{1}{r}$. That is, $V_r(r) = \frac{\mathcal{Z}}{r}$, the Coulomb potential, where $\mathcal{Z}$ is the charge coupling parameter which is proportional to the product of the charge number and the fine structure constant. It is worth noting that the radial potential function could also include a term proportional to $r^{-2}$ without destroying the solvability of Eq. (3.2). However, by a redefinition of the separation constant $E_\theta$ in Eq. (2.9c), such a term could easily be absorbed into the centripetal potential $E_\theta/r^2$. Now, for bound states $\mathcal{Z} < 0$ and the confluent hypergeometric series ${}_1F_1(p;q;y)$ must terminate which requires that $p = -k$, where $k = 0, 1, 2, \ldots$ Simple manipulations of Eq. (3.2) give the following results:

$$\tau = n + \alpha + \beta + 1,\ \xi = \tfrac{1}{2},\ q = 2\tau, \qquad (3.3a)$$

$$\varepsilon_{knm} = \mathrm{m}\frac{(k+n+\alpha+\beta+1)^2 - \tfrac{1}{4}\mathcal{Z}^2}{(k+n+\alpha+\beta+1)^2 + \tfrac{1}{4}\mathcal{Z}^2}, \qquad (3.3b)$$

$$\rho = -\mathcal{Z}\frac{\varepsilon_{knm} + \mathrm{m}}{k+n+\alpha+\beta+1}. \qquad (3.3c)$$

The dependence of the energy spectrum on the integer $m$ comes from the parameters $\alpha$ and $\beta$ as given by Eqs. (2.14). Now, if this formulation of the relativistic problem is misinterpreted then one may presume that the energy spectrum (3.3b) includes that of the relativistic Coulomb problem (when $a = b = 0$) and that of the relativistic Hartmann problem (when $a \neq 0$ and $b = 0$). Obviously, the spectra of other problems (e.g., the magnetic monopole where $\mathcal{Z} = 0$ and $b = \pm a$) could also be assumed incorrectly to be included, but will not be discussed here. For the Coulomb case $\alpha = \beta = \frac{|m|}{2}$ and the energy spectrum (3.3b) could be written as

$$\varepsilon_{k\ell} = m\frac{(k+\ell+1)^2 - \mathcal{Z}^2/4}{(k+\ell+1)^2 + \mathcal{Z}^2/4}, \tag{3.4}$$

where $\ell = n+|m| = 0,1,2,...$ This is *not* equal to the well-known positive energy spectrum of the relativistic Dirac-Coulomb problem [8],

$$\varepsilon_{k\ell} = m\left[1 + \mathcal{Z}^2\bigg/\left(k + \sqrt{(\ell+1)^2 - \mathcal{Z}^2}\right)^2\right]^{-1/2}. \tag{3.5}$$

However, both give the correct nonrelativistic limit (in the case of week coupling, i.e. $\mathcal{Z} \ll 1$ and $|E| \ll m$):

$$E_{k\ell} = -\frac{m\mathcal{Z}^2}{2(k+\ell+1)^2}. \tag{3.6}$$

As for the Hartmann problem, it is well known that taking the parameter limit $a \to 0$ gives the Coulomb problem. Therefore, we can also conclude that the correct relativistic extension of the Hartmann problem is not as formulated above [9]. Nevertheless, the non-relativistic limit ($\mathcal{Z} \ll 1$) of the energy spectrum (3.3b) for the Hartmann problem, where $a \neq 0$, $b = 0$ and $\alpha = \beta = \frac{1}{2}\sqrt{m^2 + a(\varepsilon + m)}$, is

$$E_{knm} = -\frac{m\mathcal{Z}^2}{2}\left(k + n + 1 + \sqrt{m^2 + 2ma}\right)^{-2}, \tag{3.7}$$

which is the correct nonrelativistic spectrum [6,10].

Taking $y = (\rho r)^2$ in the radial wavefunction (3.1) and employing the differential equation of the confluent hypergeometric series reduce Eq. (2.9c) to the following

$$\left[\left(1 - 2\xi + \frac{2\tau - q + 1/2}{y}\right)y\frac{d}{dy} + \frac{2\tau(2\tau-1) - \lambda(\lambda+1)}{4y} + \xi^2 y - \frac{\varepsilon + m}{4\rho^2}V_r \right.$$
$$\left. + p - \xi\left(2\tau + \tfrac{1}{2}\right) + \frac{\varepsilon^2 - m^2}{4\rho^2}\right]{}_1F_1 = 0. \tag{3.8}$$

Orthogonal representation for the bound states dictates that the factors multiplying ${}_1F_1$ and $\frac{d}{dy}{}_1F_1$ must vanish. Thus, the radial potential function should be proportional to $r^2$. That is, we can write $V_r(r) = \frac{1}{2}m\omega^2 r^2$, which is the potential for the 3D isotropic oscillator with $\omega$ being the oscillator frequency. Similar to the Coulomb problem, this radial potential could also include a term proportional to $y^{-1}$ (i.e., $r^{-2}$). Such a term could be absorbed, as well, into the centripetal potential $E_\theta/r^2$. For bound states the confluent hypergeometric series ${}_1F_1$ must terminate requiring that $p = -k$, where $k = 0,1,2,...$ Simple manipulations of Eq. (3.8) give the following results:

$$2\tau = n + \alpha + \beta + 1, \quad \xi = \tfrac{1}{2}, \quad q = 2\tau + \tfrac{1}{2}, \tag{3.9a}$$

$$(\varepsilon_{knm} - m)\sqrt{\frac{\varepsilon_{knm} + m}{2m}} = \omega\left(2k + n + \alpha + \beta + \tfrac{3}{2}\right), \tag{3.9b}$$

$$\rho^4 = \tfrac{1}{2}m\omega^2(\varepsilon_{knm} + m). \tag{3.9c}$$

The relativistic bound states energy spectrum is obtained by solving Eq. (3.9b) for $\varepsilon_{knm}$. For the spherically symmetric case, where $a = b = 0$ and $\alpha = \beta = \frac{|m|}{2}$, the right hand side of Eq. (3.9b) becomes $\omega\left(2k + \ell + \tfrac{3}{2}\right)$. The resulting formula will be compared to the well-known positive energy spectrum of the Dirac-Oscillator [11]

$$\varepsilon_{k\ell} = m \begin{cases} \sqrt{1 + 4\frac{\omega}{m}\left(k + \ell + \frac{3}{2}\right)} & ,\ell = j + \frac{1}{2} \\ \sqrt{1 + 4\omega k/m} & ,\ell = j - \frac{1}{2} \end{cases} \qquad (3.10)$$

where $j$ is the total angular momentum, orbital plus spin. It is obvious that the two relativistic spectra do not coincide. Nevertheless, the nonrelativistic limit (when $|E| \ll m$ and $\omega \ll m$) of (3.9b), which describes the oscillator in the presence of the noncentral potential (2.13), is

$$E_{knm} = \omega\left(2k + n + \alpha + \beta + \tfrac{3}{2}\right), \qquad (3.11)$$

which is the correct spectrum [7,12].

The three examples presented above (the Coulomb, Hartmann, and Oscillator) show that the formulation of the relativistic problem as depicted in Eq. (2.1a,b) [equivalently, Eq. (2.1a,b)'] with $S = +V$ should not be misinterpreted as equivalent to those that have the same nonrelativistic limit. This is compatible with the view that although the non-relativistic limit is well-defined and unique, the relativistic extension is not.

## IV. POTENTIALS WITH UNEQUAL MAGNITUDES

For the sake of completeness, we study in this section a related problem that has also received equal attention [13]. It deals with the K-G equation (2.1b) [equivalently, Eq. (2.1b)'] but with an unbalanced potential contributions in which we take $S = \eta V$, where $\eta$ is a real parameter such that $\eta \neq \pm 1$. Now, this case does not suffer from the singular potential structure mentioned below Eq. (2.1b). Additionally, the Dirac and K-G equations give results that are equivalent *only* under certain physical constraints. Thus, we expect fruitful results. As an illustrative example, we consider spherically symmetric exponential-type potentials leading to the correct formulation of the relativistic extension of the S-wave Morse problem [14]. It bears very close resemblance to the Dirac-Morse problem [15].

The radial component of the K-G equation (2.1b)' for spherically symmetric potentials, with $S = \eta V$ and $V(\vec{r}) = V_0 e^{-\rho r}$, reads as follows

$$\left[\frac{d^2}{dr^2} - \tfrac{1}{4}(\eta^2 - 1)V_0^2 e^{-2\rho r} - (\varepsilon + \eta m)V_0 e^{-\rho r} + \varepsilon^2 - m^2\right] R(r) = 0, \qquad (4.1)$$

where $V_0$ and $\rho$ are real potential parameters and $\rho$ positive. On the other hand, the Dirac-Morse potential [15] is a three-parameter relativistic extension of the S-wave Morse oscillator. The "kinetic balance" relation in that problem could be used to eliminate the lower spinor component giving the following second order radial differential equation [15] for the upper component

$$\left[\frac{d^2}{dr^2} - (A_0/\zeta)^2 e^{-2\rho r} - \left(2\varepsilon + \frac{\rho}{\zeta}\right)A_0 e^{-\rho r} + \varepsilon^2 - m^2\right] R(r) = 0, \qquad (4.2)$$

where $\{\zeta, \rho, A_0\}$ are the physical parameters of the problem such that, for bound states, $\zeta^2 = (2A_0/\rho)^2$. Comparing this equation with Eq. (4.1) shows that the current K-G problem is an S-wave Dirac-Morse problem if $A_0 = \tfrac{1}{2}V_0$, $\zeta = \tfrac{1}{2}(\rho/\eta m)$ and only for the restricted case where $\zeta^2 = (\rho/2\mathrm{m})^2 - 1$ and $\rho > 2\mathrm{m}$ (i.e., only if $\eta^2 = [1 - (2\mathrm{m}/\rho)^2]^{-1}$). To pursue the solution of the current K-G problem, we postulate the following radial wave function which is compatible with the domain of the wave operator (4.1)

$$R(r) = A z^\tau e^{-\xi z} {}_1F_1(p;q;z), \tag{4.3}$$

where $z = e^{-\rho r}$ and, for economy of notation, we used the same symbols as those in the previous section. Substituting this in Eq. (4.1) and using the differential equation of the confluent hypergeometric series we obtain the following

$$\left\{ \left(1 - 2\xi + \frac{2\tau - q + 1}{z}\right)\frac{d}{dz} + \frac{1}{z^2}\left(\tau^2 + \frac{\varepsilon^2 - \mathrm{m}^2}{\rho^2}\right) + \xi^2 - V_0^2 \frac{\eta^2 - 1}{4\rho^2} \right. \\ \left. - \frac{1}{z}\left[\xi(2\tau + 1) - p + V_0 \frac{\varepsilon + \eta\mathrm{m}}{\rho^2}\right] + \right\} {}_1F_1 = 0. \tag{4.4}$$

Real solutions of this equation dictates that $\eta V_0 < 0$. For bound states we must also have $p = -n$, where $n = 0, 1, 2, \ldots$ Moreover, the wavefunction parameters are evaluated as follows

$$\tau = \chi - n - \tfrac{1}{2} - (V_0/\rho^2)\varepsilon_n, \quad \xi = \tfrac{1}{2} \tag{4.5a}$$

$$q = 2\tau + 1, \quad \eta^2 = 1 + (\rho/V_0)^2, \tag{4.5b}$$

and the relativistic energy spectrum is derived as

$$\left[1 + (V_0/\rho)^2\right]\varepsilon_n = V_0\left(\chi - n - \tfrac{1}{2}\right) \pm \rho\sqrt{\left(n + \tfrac{1}{2}\right)\left(2\chi - n - \tfrac{1}{2}\right)}, \tag{4.6}$$

where $\chi = \frac{\mathrm{m}}{\rho}\sqrt{1 + (V_0/\rho)^2}$ and $n \leq 2\chi - \tfrac{1}{2}$. These results show that the Klein-Gordon-Morse problem is equivalent to the Dirac-Morse problem if and only if the potential parameters are related as $V_0^2 = \rho^2[(\rho/2\mathrm{m})^2 - 1]$.

## V. CONCLUDING REMARKS

In this paper, we studied the three dimensional Dirac and Klein-Gordon equations with scalar $S$ and vector $V$ potentials such that $S = \pm V$. This type of coupling attracted a lot of attention in the literature due to the resulting simplification in the solution of the relativistic problem. The wave equation could always be reduced to a Schrödinger-type second order differential equation. This puts at one's disposal a variety of well established analytic tools and techniques to be employed in the analysis and solution of the problem. These techniques have been well developed over the years by many researchers in dealing with the Sturm-Liouville problem and the Schrödinger equation. However, we showed that the nonrelativistic limit of the case $S = -V$ results in a trivial non-interacting theory. As for the case $S = +V$ few general remarks were given. These include the singular structure of the potential, the missing negative energy spectrum, and the unexpected close affinity of the Klein-Gordon equation to the Dirac equation in the presence of interaction. Several illustrative examples with $S = +V$ were considered in order to support our conclusions and give a clear presentation of our study. We considered the more general situation where the potentials are not only radial but non-central. The relativistic Coulomb, Oscillator and Hartmann potentials were considered. We found that the resulting relativistic energy spectrum does not agree with those of well-known relativistic extensions that have the same nonrelativistic limit. This shows that although the nonrelativistic limit is unique, the relativistic extension is not.

Finally, we looked at the case where $S = \eta V$ such that $\eta \neq \pm 1$. This results in uneven contribution of the two potentials. Moreover, the structure of the interaction is no longer singular. We studied the spherically symmetric problem with radial exponential potentials and showed that an acceptable extension of the S-wave Morse problem is obtained. Bound states exist under certain conditions and the relativistic energy spectrum was derived.

## ACKNOWLEDGMENTS

ADA and HB acknowledge the support of King Fahd University of Petroleum and Minerals. AA is grateful to the Physics Department at King Faisal University for support. Motivating discussions with M. S. Abdelmonem are highly appreciated.